\documentclass[%
preprint,
superscriptaddress,
amsmath,amssymb,
prx,
]{revtex4-2}
\usepackage{color}
\usepackage{xcolor}
\usepackage{soul}
\usepackage{float}

\usepackage[pdftex]{graphicx}
\usepackage{graphicx}
\usepackage{dcolumn}
\usepackage{bm}
\usepackage{siunitx}
\usepackage{dirtytalk}
\usepackage{comment}
\usepackage{fixltx2e}
\usepackage{soul}
\usepackage{svg}
\usepackage{braket}
\usepackage{lineno}

\usepackage{hyperref}
\hypersetup{
    bookmarks=true,         
    unicode=false,          
    pdftoolbar=true,        
    pdfmenubar=true,        
    pdffitwindow=false,     
    pdftitle={Dispersive readout of an isolated 2x2 quantum dot array in silicon},    
    pdfauthor={Dr. Matias Urdampilleta},     
    pdfsubject={},   
    pdfcreator={Dr. Matias Urdampilleta},   
    pdfproducer={},  
    pdfkeywords={,} {} {}, 
    pdfnewwindow=true,      
    colorlinks=false,       
    linkcolor=red,          
    citecolor=green,        
    filecolor=magenta,      
    urlcolor=cyan           
}

\usepackage{filecontents}
\usepackage{comment}

\DeclareUnicodeCharacter{2009}{\,} 
\DeclareSIUnit{\belmilliwatt}{Bm}
\DeclareSIUnit{\dBm}{\deci\belmilliwatt}

\begin{document}
\bibliographystyle{apsrev4-1}
\setstcolor{red}

\preprint{}
\title{Combining multiplexed gate-based readout and isolated CMOS quantum dot arrays
}

\author{Pierre Hamonic}
\affiliation{Univ. Grenoble Alpes, CNRS, Grenoble INP, Institut N\'eel, 38402 Grenoble, France}

\author{Martin Nurizzo}
\affiliation{Univ. Grenoble Alpes, CNRS, Grenoble INP, Institut N\'eel, 38402 Grenoble, France}

\author{Jayshankar Nath}
\affiliation{Quobly, Grenoble, France}

\author{Matthieu C. Dartiailh}
\affiliation{Quobly, Grenoble, France}

\author{Victor El-Homsy}
\affiliation{Univ. Grenoble Alpes, CNRS, Grenoble INP, Institut N\'eel, 38402 Grenoble, France}

\author{Mathis Fragnol}
\affiliation{Univ. Grenoble Alpes, CNRS, Grenoble INP, Institut N\'eel, 38402 Grenoble, France}

\author{Biel Martinez}
\affiliation{Univ. Grenoble Alpes, CEA, Leti, F-38000 Grenoble, France}

\author{Pierre-Louis Julliard}
\affiliation{Quobly, Grenoble, France}

\author{Bruna Cardoso Paz}
\affiliation{Quobly, Grenoble, France}

\author{Mathilde Ouvrier-Buffet}
\affiliation{Univ. Grenoble Alpes, CNRS, Grenoble INP, Institut N\'eel, 38402 Grenoble, France}

\author{Jean-Baptiste Filippini}
\affiliation{Univ. Grenoble Alpes, CNRS, Grenoble INP, Institut N\'eel, 38402 Grenoble, France}

\author{Benoit Bertrand}
\affiliation{Univ. Grenoble Alpes, CEA, Leti, F-38000 Grenoble, France}

\author{Heimanu Niebojewski}
\affiliation{Univ. Grenoble Alpes, CEA, Leti, F-38000 Grenoble, France}

\author{Christopher B{\"a}uerle}
\affiliation{Univ. Grenoble Alpes, CNRS, Grenoble INP, Institut N\'eel, 38402 Grenoble, France}

\author{Maud Vinet}
\affiliation{Quobly, Grenoble, France}

\author{Franck Balestro}
\affiliation{Univ. Grenoble Alpes, CNRS, Grenoble INP, Institut N\'eel, 38402 Grenoble, France}

\author{Tristan Meunier}
\affiliation{Univ. Grenoble Alpes, CNRS, Grenoble INP, Institut N\'eel, 38402 Grenoble, France}
\affiliation{Quobly, Grenoble, France}

\author{Matias Urdampilleta	}
\email{matias.urdampilleta@neel.cnrs.fr}
\affiliation{Univ. Grenoble Alpes, CNRS, Grenoble INP, Institut N\'eel, 38402 Grenoble, France}

\date{\today}

\begin{abstract}
Semiconductor quantum dot arrays are a promising platform to perform spin-based error-corrected quantum computation with large numbers of qubits. 
However, due to the diverging number of possible charge configurations combined with the limited sensitivity of large-footprint charge sensors, achieving single-spin occupancy in each dot in a growing quantum dot array is exceedingly complex.
Therefore, to scale-up a spin-based architecture we must change how individual charges are readout and controlled.
Here, we demonstrate single-spin occupancy of each dot in a foundry-fabricated array by combining two methods. 1/ Loading a finite number of electrons into the quantum dot array; simplifying electrostatic tuning by isolating the array from the reservoirs. 
2/ Deploying multiplex gate-based reflectometry to dispersively probe charge tunneling and spin states without charge sensors or reservoirs.
Our isolated arrays probed by embedded multiplex readout  can be readily electrostatically tuned. They are thus a viable, scalable approach for spin-based quantum architectures.

\end{abstract}

\maketitle

\section{INTRODUCTION}

Semiconductor spin qubit arrays hold significant promise for the future of quantum computing, primarily due to their easier scalability thanks to existing semiconductor industry processes  \cite{Gonzalez-Zalba2021, PhysRevApplied.6.054013, Veldhorst2017, Patomaki2024, Crawford2023}.
By facilitating the integration of a larger number of qubits \cite{Mortemousque2021, Borsoi2023, Philips2022, Thorvaldson2024, Weinstein2023}, semiconductor spin qubit arrays should enable complex computations and enhance problem-solving capabilities. 
In this context, developing semiconducor spin qubits using industrially compatible process flows should lead to reliable devices produced at high yields with low variability \cite{Zwerver2022, Neyens2024}, cointegration with cryoelectronics is also possible \cite{Bartee2024}. Furthermore, the redundancy potential of large arrays is compatible with the implementation of quantum error correction techniques \cite{Takeda2022}, which are a critical component if we are to achieve fault-tolerant quantum computation.

However, the control and readout of large, dense arrays  still presents a number of challenges.
Firstly, precise tuning of charge configurations faces various technical hurdles, such as handling capacitive coupling between neighboring quantum dots \cite{Mills2019, Borsoi2023, Mortemousque2021, Unseld2023} or dealing with a diverging number of accessible charge configurations in arrays open to reservoirs \cite{Heinz2021, Lawrie2023, Eoin2023}. 
Secondly, readout of qubit states must be consistent if we are to extract useful information from the quantum system. 
The charge detectors currently used for readout from arrays often have a large footprint and limited depth of sensitivity  \cite{Blumoff2022, Philips2022, Unseld2023, Borsoi2023, Mortemousque2021, Oakes2023, Niegemann2022, Ansaloni2020}. Consequently, they are incompatible with the control of dense arrays.
To address these challenges, innovative strategies must be developed to distribute and control single charges in an array and to develop new measurement schemes. 

In this paper, we describe a novel approach which consists in operating a small array as an isolated unit cell with embedded charge and spin readout. 
In this regime, the number of accessible charge configurations is relatively small compared to arrays open to reservoirs. This reduced complexity greatly facilitates electrostatic tuning \cite{Nurizzo2022,Bertrand2015, Eenink2019, Yang2020, PRXQuantum.2.030331}. 
We exploit gate-based reflectometry to achieve embedded charge and spin readout. This method is potentially scalable if used to probe interdot charge transitions and does not require a dedicated charge detector \cite{Ruffino2022, Veldhorst2017}. Moreover, operating in the dispersive regime (where drive frequency is small compared to tunnel couplings) allows for a clear identification of spin states using quantum capacitance spectroscopy \cite{PhysRevApplied.16.034031}: a singlet state gives finite quantum capacitance at the charge transition, changing the resonator's response; and a triplet state leaves the response unchanged \cite{Petersson2010, Lundberg2020, Landig2019, Crippa2019, Cottet2011}.
The results presented here provide guidelines for scaling-up by replicating the unit cell to build larger bilinear or 2D qubit arrays, and pave the way for the design of spin-based quantum architecture.

\section{Load, isolate and read a double quantum dot}

\begin{figure}%
\includegraphics[width=\columnwidth]{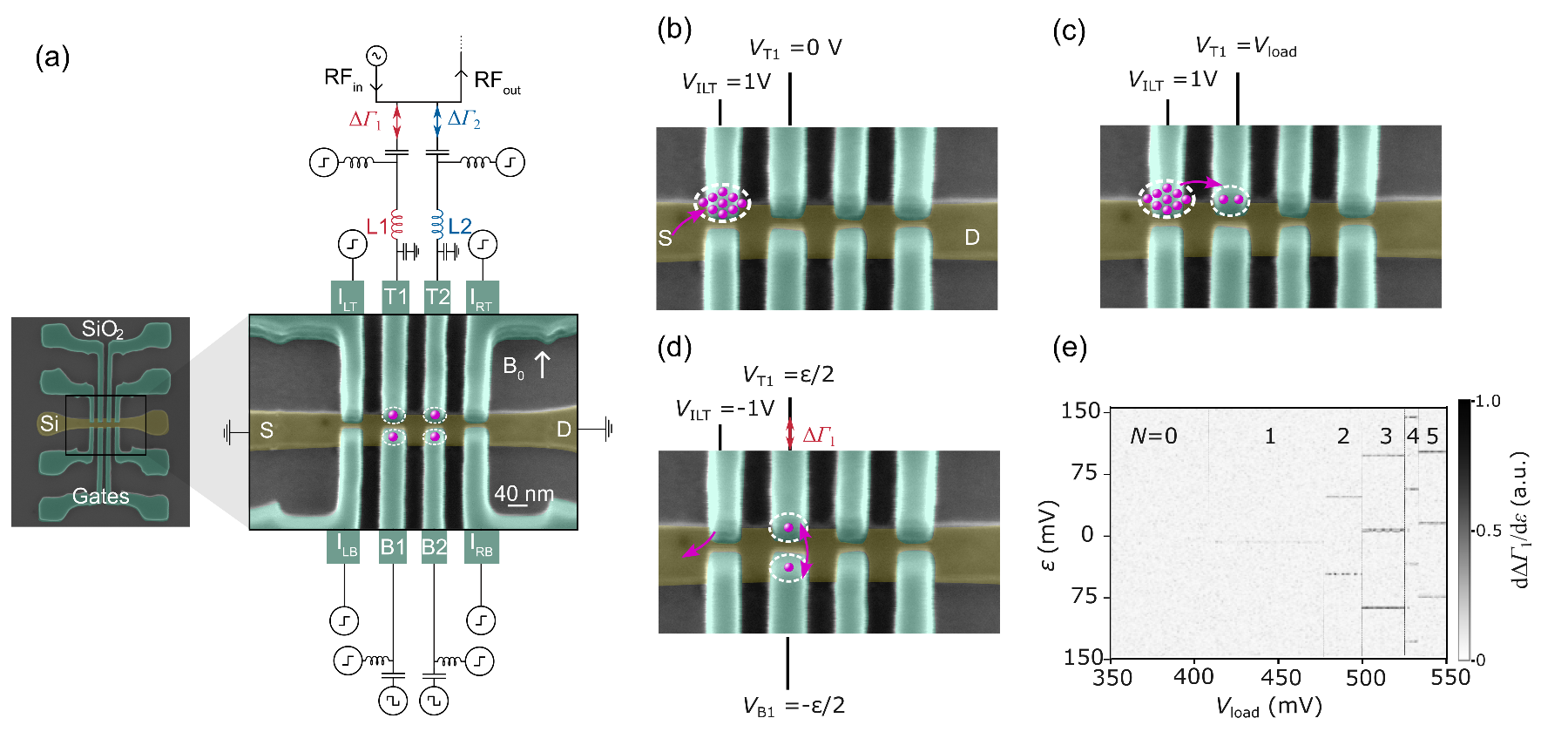}
\caption{\textbf{Device description and double quantum dot stability diagram} 
\textbf{(a)} SEM micrograph of the 4 split-gate devices. The source (S) and drain (D) ends of the channel are doped, forming electron reservoirs.
On the right, an electronic diagram of the experiment is shown. T1 and T2 are connected to tank circuits comprising L\textsubscript{1} and L\textsubscript{2} and the capacitance to ground.
\textbf{(b)} First, a large voltage (1 V) is applied to $I_{LT}$ to accumulate electrons in the underlying QD.
\textbf{(c)} Then, a finite voltage V\textsubscript{load} is applied to T1 to transfer a given number of electrons to the underlying QD. 
\textbf{(d)} Finally, $I_{LT}$ is pulsed back to 0 V. At this point, the electrons are trapped in the central array and they can be probed by reflectometry by sweeping the detuning $\epsilon_{X}$ between T1 and B1.
\textbf{(e)} Derivative of the resonator response at $f_1$ as a function of double dot detuning after completion of the loading procedure. $V_{load}$ corresponds to the gate voltage applied to T1 during the loading sequence. Horizontal lines indicate charge tunneling between the two dots.
The number of lines for a given $V_{load}$ corresponds to the number of electrons in the isolated structure.
}
\label{fig:device}%
\end{figure}

Our device (Figure \ref{fig:device}(a)) was fabricated in an industrial-research foundry using 300-mm CMOS processes on a silicon-on-insulator substrate. 
It features a 40 nm-wide 10 nm-thick silicon channel, separated from the substrate by a 145-nm buried oxide layer. 
Titanium nitride and poly-silicon gates are isolated from the nanowire by a 6-nm layer of thermally grown silicon dioxide. 
A combination of deep-ultraviolet and electron-beam lithography is used to pattern the gate structure, as described in references \cite{Niegemann2022, Klemt2023}. 

The 8 gates of the device, arranged in a 2x4 pattern, are labeled as follows: external gates labeled isolation gates ($I_{LB}$, $I_{RB}$, $I_{RT}$ and $I_{LT}$), and central gates, which form the quantum dot array, labeled either top (T1, T2) or bottom (B1, B2) (fig.\ref{fig:device}(b)).
All gates are 40 nm wide and spaced 40 nm apart in both longitudinal and transversal directions. These dimensions were selected to have nominal transverse tunnel couplings in the range of a few GHz, since the dispersive regime requires tunnel coupling to be relatively large compared to RF drive frequency and amplitude \cite{Vigneau2023, GZ2015}. 
Silicon nitride spacers (35 nm thick) were deposited between the gates before doping the source and drain reservoirs to implant ions. 
The whole device is encapsulated, and the gates are connected to aluminum bond pads through standard Cu-damascene back-end-of-line processing.

The device is electrically operated as follows: the gates are DC-biased, and T1, T2, B1 and B2 are also connected to a bias tee with cutoff frequency $f_c = \SI{30}{\kilo\hertz}$. In addition, T1 and T2 are connected to two tank circuits (fig \ref{fig:device}(a)),
each comprising an Nb spiral inductor. Inductance was set to $L\textsubscript{1} = \SI{69}{\nano\henry}$ for gate T1 and $L\textsubscript{2} = \SI{120}{\nano\henry}$ for gate T2. 
When combined with the parasitic capacitance, C\textsubscript{P}, of the device, the final LC resonance frequencies were $f\textsubscript{1} = \SI{1.2}{\giga\hertz}$ and $f\textsubscript{2} = \SI{0.8}{\giga\hertz}$. 
At zero magnetic field, $\text{C}\textsubscript{P} = \SI{0.25}{\pico\farad}$ is extracted from the resonance frequencies, with quality factors of around 50 and 20, respectively. 
The amplitude variation of the reflected radio frequency signal close to the resonance (noted $\Delta\Gamma$ in Fig. \ref{fig:device}(a)) is determined using analog demodulation.
At the base temperature of the dilution fridge, applying a positive voltage to the gates leads to the accumulation of quantum dots at the Si-SiO\textsubscript{2} interface, allowing the formation of a 2×2 quantum dot array when the isolation gates I$_{LB}$, I$_{RB}$, I$_{RT}$ and I$_{LT}$ are used as barriers with respect to the reservoirs.

To isolate charges in the 2x2 central array, we first completely deplete the device by applying negative voltages to all  gates.
Next, we open I$_{LT}$ to accumulate a reservoir underneath (Fig. \ref{fig:device}(c)) and bias T1, the loading dot, at a finite voltage noted $V_{load}$ (Fig. \ref{fig:device}(d)).
By rapidly negatively pulsing I$_{LT}$, we isolate electrons in dot T1 (Fig. \ref{fig:device}(e)). 
We then perform quantum capacitance measurements at $f_1$ in the isolated double quantum dot (DQD) T1-B1 and plot them against the voltage applied to T1 during the loading sequence (Figure \ref{fig:device}(f)).
In this representation, the horizontal lines correspond to interdot charge transitions (ICT) in the isolated DQD T1-B1.
The number of lines for a given loading voltage corresponds to the number of electrons in the isolated structure.
This is further verified by combining traditional charge detection and quantum capacitance measurements. A very good correlation was obtained for the additions of the first electrons (see Supplementary Materials). 
In subsequent experiments, $V_{load}$ applied during the loading sequence can be adjusted as a function of how many electrons we want to work with.
Depending on the voltage detuning between the isolation and plunger gates, the electrons loaded can be trapped in the structure for minutes to weeks. 
Hereafter, we biased our isolation gates to ensure that electrons remained trapped for at least three hours.

\section{Operate an isolated triple quantum dot}
\begin{figure}%
\includegraphics[width=\columnwidth]{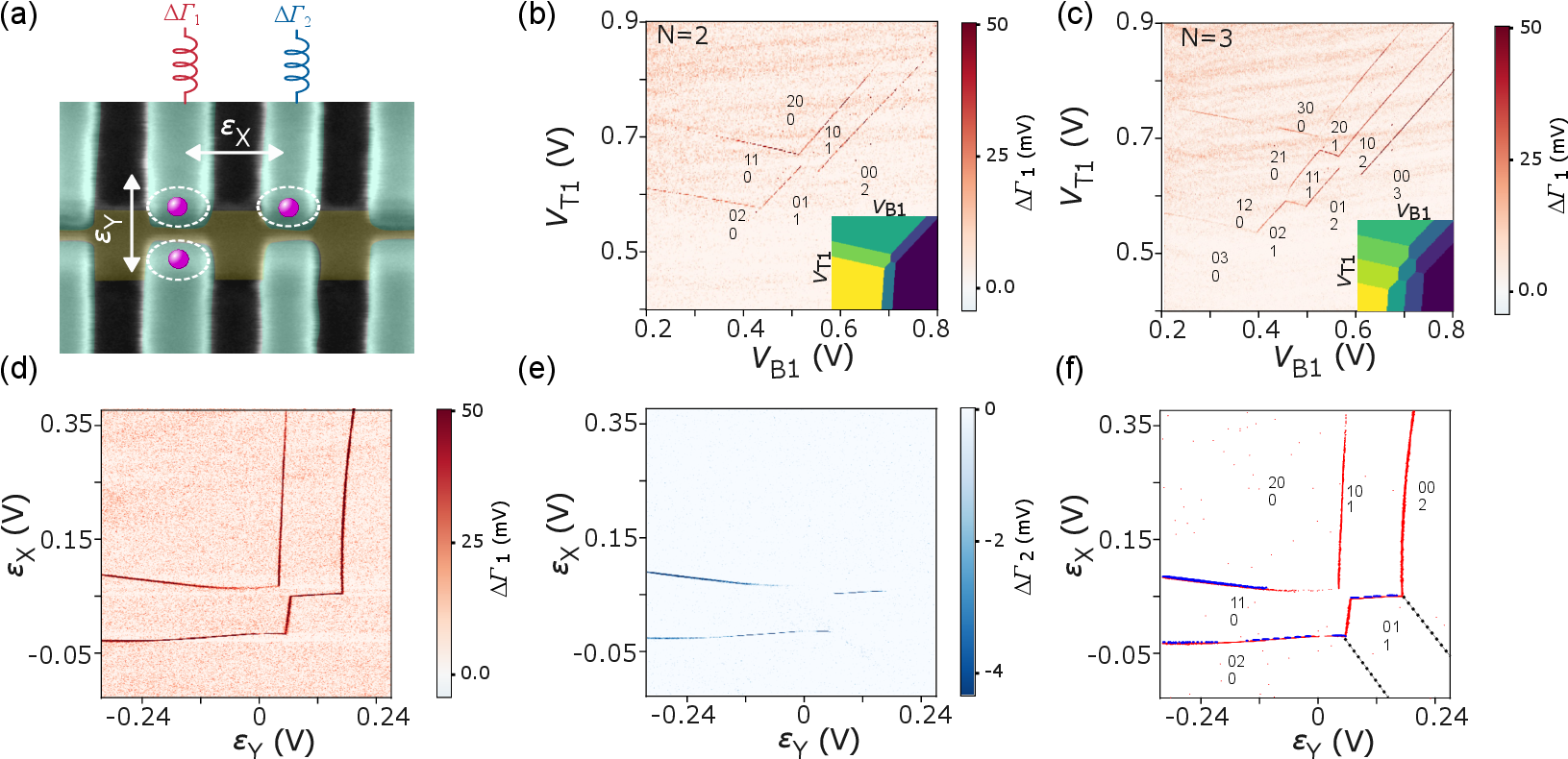}
\caption{\textbf{The isolated triple quantum dot} 
\textbf{(a)} SEM micrograph of the central array, showing the isolated TQD and the two detuning axes involved. 
\textbf{(b),(c)} Stability diagrams for the TQD loaded with 2 and 3 electrons, respectively. The insets show electrostatic simulations of the stability diagrams. 
\textbf{(d,e)} Stability diagrams using virtual gates for the 2-electron TQD probed at $f_1$ and $f_2$ respectively, in frequency multiplexing. 
\textbf{(f)} Sum of the two threshold signals extracted from (d) and (e). 
}
\label{fig:TQD}%
\end{figure}

Having demonstrated electron loading and control in a DQD, we extended tuning to a triple quantum dot (TQD).
A right angle triangular TQD unit cell (see Fig. \ref{fig:TQD}(a)) is coupled along two orthogonal directions, paving the way for a 2D qubit lattice.
We started the characterization by loading N electrons into the storage dot T1.  
Leaving gate T2 at V\textsubscript{T2}=0.6 V, we mapped the stability diagram for gates T1 and B1 with resonator 1, producing Fig \ref{fig:TQD} (b) and (c) for N=2 and N=3, respectively. 
These figures show two sets of lines: one set with slope close to unity and one set with a slightly negative slope.
The first set corresponds to ICTs between T1 and B1 and the second set to ICTs between T1 and T2,
as confirmed by the electrostatic simulations (inset on each graph).
Due to insignificant tunnel coupling, the vertical lines visible in the simulations, corresponding to ICTs between T2 and B1, cannot be resolved in the experimental stability diagram.

A TQD configuration of particular relevance for quantum information processing is the $\bigl(\begin{smallmatrix}
1&1\\ 1&
\end{smallmatrix} \bigr)$ charge occupation state \cite{Weinstein2023}, which is achieved in the middle of the stability diagram for 3 electrons  (Fig. \ref{fig:TQD}(c)).
This charge state can be readily induced starting with the DQD which gives the best signal (T1-B1) and tuning it to the (2,1) state before reducing T1 to transfer one electron to T2. Applying this procedure produces unambiguous, clearly visible transitions.
Moreover, the $\bigl(\begin{smallmatrix}
1&1\\ 1&
\end{smallmatrix} \bigr)$ charge occupation state is further confirmed by spin and valley measurements at different ICTs closing this charge state (see Supplementary Materials).

For arrays with capacitive cross couplings, virtual gates are required to achieve accurate tuning  \cite{Philips2022, Mills2019, Unseld2023, Borsoi2023, Mortemousque2021}.
In our array, we extracted a lever arm ratio matrix from the stability diagram and simulations \cite{Flentje2017}. Assuming constant electrostatic interactions we constructed the virtual gates matrix.
Fig. \ref{fig:TQD}(d) presents a 2-electron TQD stability diagram for the two detuning axes (T1-T2 and T1-B1) along the two arms of the TQD folowing application of virtual gates.
The ICTs are now mostly aligned with the horizontal ($\epsilon$\textsubscript{x}) and vertical ($\epsilon$\textsubscript{y}) detuning axis directions, confirming the good approximation of our lever arm ratios and of the constant interaction model.

Frequency multiplexing can provide additional information about which QD is involved in the different ICTs. This information is required to explore a larger number of quantum dots in interaction.
As an illustration, we start here with the simple TQD case but will expand to a 2x2 array in the next section.
Figures \ref{fig:TQD}(d) and (e) show the same stability diagram probed at the same time but at two different frequencies.
At $f_1$, we see the transitions involving T1, while at $f_2$ we see those involving T2. 
Therefore, when we add these two signals (see Fig. \ref{fig:TQD}(f)) we can confirm that the horizontal transitions (blue and red) correspond to an electron tunneling between T1 and T2, whereas the vertical transitions (red only) involve T1 and B1.

The combination of virtual gate and frequency multiplexing greatly facilitates interpretation of the stability diagram, and will be further applied to the 2x2 array.

\section{Multiplexed readout of the 2x2 array}

\begin{figure}%
\includegraphics[width=\columnwidth]{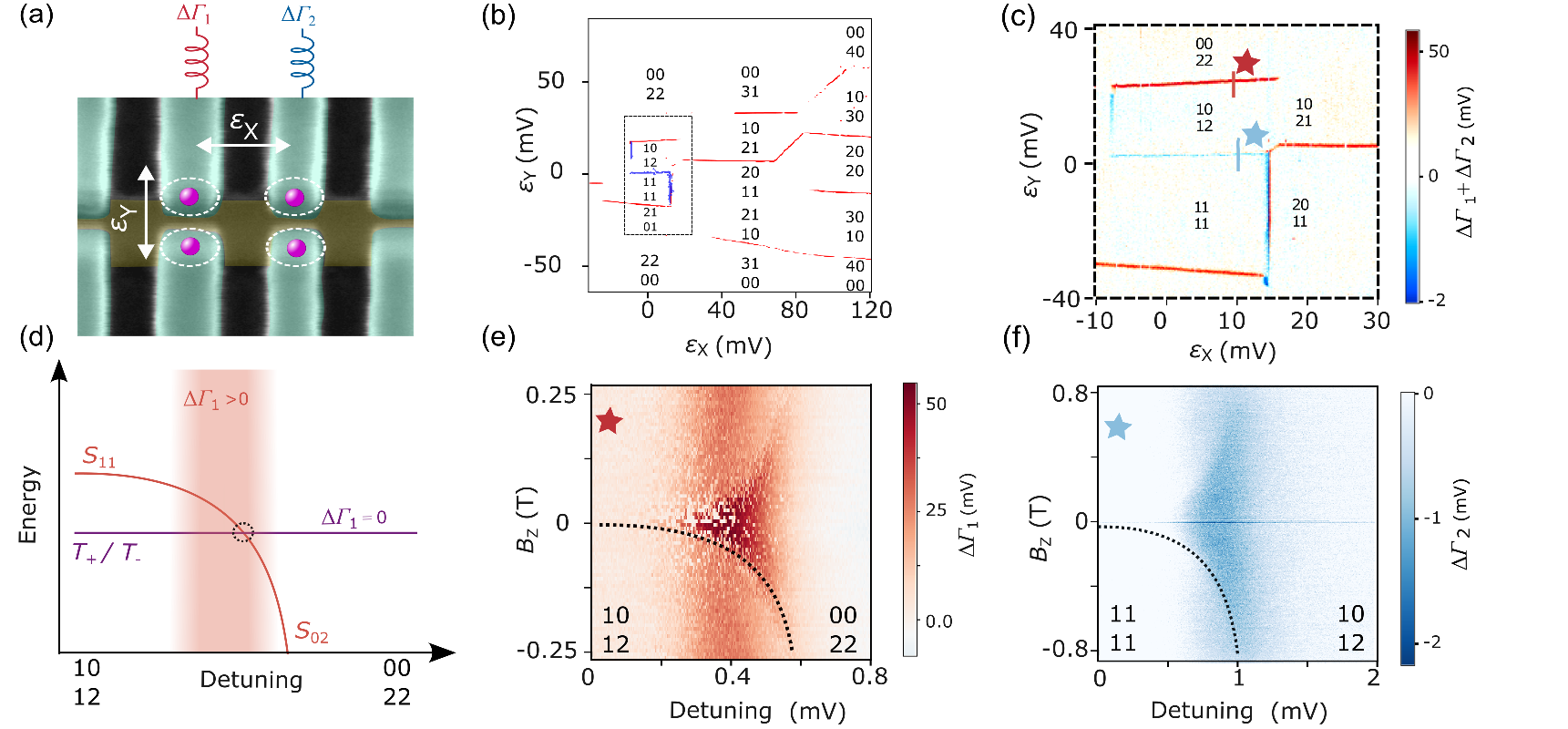}
\caption{\textbf{Isolated 2x2, 4-electron array.} 
\textbf{(a)} SEM micrograph of the central 2x2 array.
\textbf{(b)} Stability diagram plotting vertical and horizontal detuning. 
At large horizontal and vertical detuning, the 4 electrons are localized in B1 ($\bigl(\begin{smallmatrix}
0&0\\ 4&0
\end{smallmatrix} \bigr)$).
As the horizontal detuning is reduced below 30 mV, two electrons are transferred to the right column.
\textbf{(c)} Zoom of the previous stability diagram using combined signal from the two multiplexed frequencies.
At positive vertical detuning, the 4 electrons are present in the bottom row in the $\bigl(\begin{smallmatrix}
0&0\\ 2&2
\end{smallmatrix} \bigr)$ charge state. As the detuning is reduced to zero, one electron per column tunnels to the top row to produce the $\bigl(\begin{smallmatrix}
1&1\\ 1&1
\end{smallmatrix} \bigr)$ charge state.
\textbf{(d)} Energy diagram of the two ground spin states for the left column during the $\bigl(\begin{smallmatrix}
0&0\\ 2&2
\end{smallmatrix} \bigr)$ to $\bigl(\begin{smallmatrix}
1&0\\ 1&2
\end{smallmatrix} \bigr)$ transition.
\textbf{(e,f)} Quantum capacitance magneto-spectroscopy of the two ICTs labeled with stars in (c). Each plot shows a spin-funnel feature characteristic of a transition involving two spins in two QDs. The dashed line is a parabolic fit showing the position of the ST- (ST+) crossing the applied field.
}
\label{fig:QQD}%
\end{figure}

Having demonstrated control and multiplexed readout of the TQD, we now move to the tuning of the 2x2 quantum dot array in the single-electron-occupancy regime.
After extracting the virtual gate matrix, linear combinations of virtual gates are built corresponding to vertical ($\epsilon$\textsubscript{y} between the lines) and horizontal ($\epsilon$\textsubscript{x} between the columns) detuning of the 2x2 array.
As a result, and like with the TQD configuration, any ICT detected can be labeled as a transition between rows if it produces a horizontal ICT, or between columns if it produces a vertical ICT (tunneling occurs along the diagonal). 
The isolated regime can be used to tune the array by applying a simple strategy. For instance, we start with 4 electrons in the the left column. Using horizontal detuning, two electrons are transferred to the right column. Then, vertical detuning is used to reach the point where one electron occupies each state.
More precisely, we start by applying large negative horizontal detuning to the 4 electrons residing in the DQD T1-B1 (left column in the array).
Four horizontal ICTs are observed at $\epsilon_X = 120mV$, corresponding to tunneling of the four electrons between T1 and B1 (Fig. \ref{fig:QQD}(b)).
As $\epsilon_X$ is reduced to zero, two electrons have tunneled to the right column. For $\epsilon_X=0$ and $\epsilon_Y=0$, there is one electron in each dot of the left column. 
We now take advantage of the multiplexed readout to determine how electrons are distributed in the second column around zero detuning.
Figure\ref{fig:QQD}(c) presents the multiplexed signal around zero detuning. Like in Fig. \ref{fig:TQD}, the combination of color signature and direction of transitions allows us to label the transitions around zero detuning.

Although only one line is visible in the second column, we further confirm that we have indeed identified the $\bigl(\begin{smallmatrix}
1&1\\ 1&1
\end{smallmatrix} \bigr)$ state by analysing quantum capacitance magnetospectroscopy during the $\bigl(\begin{smallmatrix}
1&1\\ 1&1
\end{smallmatrix} \bigr)$ to $\bigl(\begin{smallmatrix}
1&0\\ 1&2
\end{smallmatrix} \bigr)$ and the
$\bigl(\begin{smallmatrix}
1&0\\ 1&2
\end{smallmatrix} \bigr)$ to $\bigl(\begin{smallmatrix}
0&0\\ 2&2
\end{smallmatrix} \bigr)$ transitions, using resonator 2 and 1, respectively.
Assuming each column starts in the (1,1) configuration, these two electrons can either form a singlet or a triplet state, the difference in population between these two states depends on the magnetic field and on the temperature during detuning.
Moreover, as the detuning is swept across the singlet triplet anti-crossing (see Fig. \ref{fig:QQD}(d)), a triplet ground state can become an excited state and vice versa.
Knowing that only the singlet state gives a finite quantum capacitance response at the ICT, tracking the reflected signal with magnetic field provides information on the spin states present.
Figure \ref{fig:QQD}(e) and (f) show a funnel-like quantum capacitance variation as a function of the magnetic field, which is consistent with a singlet-triplet minus ST- response: at zero field, the ground singlet state produces a finite quantum capacitance signal at the ICT, while as the magnetic field increases, the zero quantum capacitance triplet state becomes the ground state. 
The fluctuations in the background are due to thermal population of the singlet states.
These data can be modeled with a two-spin toy model, where tunnel couplings of $\SI{40}{\micro\electronvolt}$ and $\SI{200}{\micro\electronvolt}$ are extracted, respectively, for the left and right columns.
These spin signatures confirm that these two transitions are appropriately labeled and that probing quantum capacitance is a viable method to measure spin qubits in the array.

\section{Conclusion}
In conclusion, we have operated a quantum dot array in the isolated regime. Charge occupancy states are characterized by gate-based reflectometry, without involving a charge sensor. Moreover, the low static disorder makes it possible to form clean double, triple, and quadruple quantum dots using virtual gates. Single-occupancy can be demonstrated in all configurations by probing quantum capacitances at the interdot charge transitions.
Reading an isolated array with gate-based reflectometry thus offers easy electrostatic tuning and a scalable approach to control larger arrays.
\section{METHODS}

Experiments were performed on the CMOS device shown in Fig. \ref{fig:device}(a) using a dilution refrigerator with a base temperature of 30 mK. Gate-based reflectometry was achieved through analog modulation/demodulation followed by digitization using an NI ADC board. The per-point acquisition time is \SI{100}{\micro\second}.
Gate voltages were applied using digital-to-analog converters controlled by an sbRIO-9208 FPGA board.

\section{ACKNOWLEDGEMENTS}
We acknowledge technical support from L. Hutin, D. Lepoittevin, I. Pheng, T. Crozes, L. Del Rey, D. Dufeu, J. Jarreau, C. Hoarau and C. Guttin. 
We thank E. Chanrion, P.-A. Mortemousque, V. Champain and B. Brun-Barriere for fruitful discussions.
This work was supported by the France 2030 program through the ANR-22-PETQ-0002 project. 
This project was also funded through the QuCube project (Grant agreement No.810504) and the QLSI consortium (Grant agreement No.101174557).
We also thank  the MOSquito project for initiating device fabrication.

\section{AUTHOR CONTRIBUTIONS}
P.H. and M.N. carried out the experiment with the help of J.N. 
M.C.D. wrote the instrumental interface environment to control the setup. 
B.B and H.N. fabricated the device. 
B.M. and P.-L.J. simulated the stability diagram and the electrostatics of the device.  
M.U. supervised the project with the help of T.M. 
M.U. wrote the manuscript in consultation with all the authors.

\section{COMPETING INTERESTS}
The authors declare no competing financial or non-financial interests

\section{DATA AVAILABILITY}
All data that support the findings of this study are available from the corresponding author upon request.

\section{Appendix}
\appendix

\section{Loading sequence and calibration of the number of electrons loaded}
\begin{figure}%
\includegraphics[width=\columnwidth]{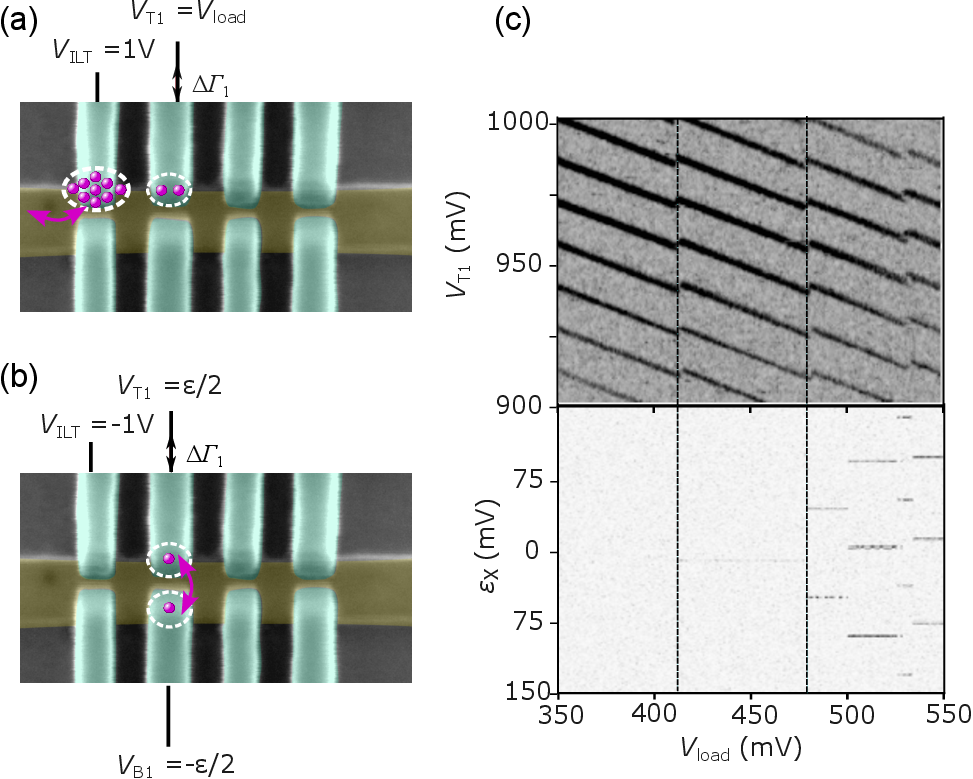}
\caption{\textbf{Calibration of the number of charge injected in the array} 
\textbf{(a)} We use the dot under T1 in the many-electron regime for charge sensing.
\textbf{(b)} The electrons in T2 are distributed in the isolated T2 B2 DQD.
\textbf{(c)}  Comparison of charge detection method and dispersive probing of ICTs. The charge sensing is achieved by using a single electron box under $I_{LT}$. The corresponding Coulomb peaks are probed using the resonator plugged on T1 through capacitive coupling.The comparison shows almost perfect correlation for the first two electrons. Beyond two electrons, the correlation fails, likely due to electrons tunneling from T1 to T2 during isolation of the array.
}
\label{fig:loading_seq2}%
\end{figure}

The loading sequence is the first step to be calibrated when operating an isolated quantum dot array.
It is important to start with the right number of electrons, which will subsequently be deployed within the array.
The main difficulty resides in the absence of absolute charge sensor, replaced by a differential charge detection method. Consequently, the dispersively probed quantum capacitance requires relatively large tunnel coupling to produce a measurable signal, making the first charges with poor tunnel coupling difficult to detect.
However, iff we have detected ICTs evolving regularly with $V_{load}$, we then correlate the loading of the first electron using a more traditional charge detection method:
During step (b) of the loading sequence, we monitor the signal reflected by the tank circuit connected to T2 while increasing $V_{load}$. In this configuration, dot T1 is used as a detector dot. It is used to perform charge detection and determine the number of charges loaded into the storage dot (under T2).
Figure \ref{fig:loading_seq2} presents these measurements with the corresponding dot configuration. The top part of Fig. \ref{fig:loading_seq2}(c) presents the signal reflected by tank circuit T2 with dot T1 operating as a charge detector \cite{Oakes2023, Niegemann2022, Ansaloni2020}. As $V_{load}$ increases, we detect charges being loaded into T2.
We then look at correlation between charge detection and dispersive measurement. The first electron is detected at the same $V_{load}$ as the first ICT,
and respectively for the second electron. However, beyond two electrons, the correlations fail.
It is likely that when the storage dot T2 contains a large number of electrons the tunnel coupling with the detector T1 increases. As a result, when the T1 gate is abruptly depleted, some of the T1 electrons tunnel into T2 instead of going to the reservoir.
With small numbers of electrons in T2, the tunnel coupling is relatively small compared to the T1-reservoir coupling. Therefore, all T1 electrons end up in the reservoir.
This is not an issue as long as there is perfect correlation for the first electron and the ICT increments regularly with $V_{load}$.

\section{Spin and valley physics in the TQD}

 \begin{figure}%
\includegraphics[width=\columnwidth]{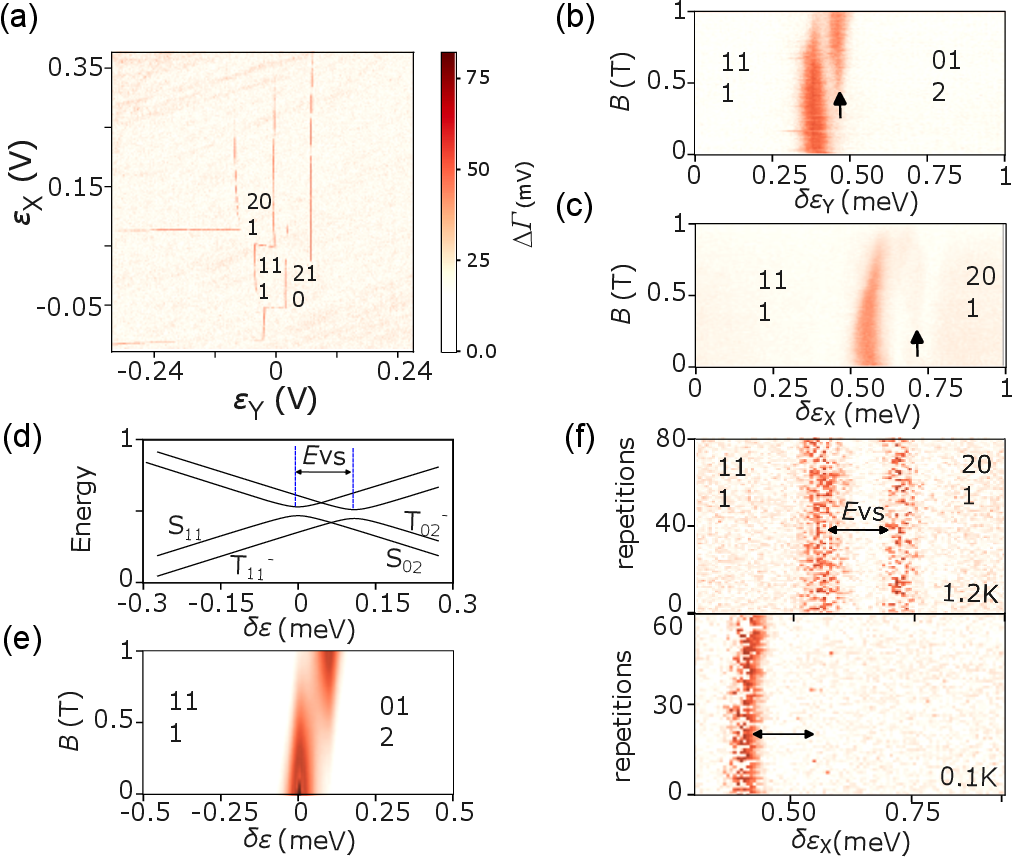}
\caption{\textbf{Spins and valleys in the TQD} 
\textbf{(a)} Stability diagram of the T1, T2, B1 isolated triple quantum dot. The  small red and blue lines indicate the detuning slices explored by magnetoscpectroscopy  
\textbf{(b) and (c)}. The dispersive signal is measured around the ICT as a function of magnetic field. The signature of a low-lying valley state is highlighted by the black arrows.
\textbf{(d)} Simplified energy diagram for the singlet and triplet states in the presence of a low-lying valley state. The electron can tunnel from 11 to 02 at two positions, one involves singlets only and one involves triplets only.
\textbf{(e)} Simulation of the quantum capacitance response around the ICT using the simplified diagram presented in (d). The pattern accurately reproduced the data in (b), and a valley splitting at around \SI{100}{\micro\electronvolt} can be extracted.
\textbf{(f)} Dispersive measurement across the ICT at different temperatures. The vertical axis corresponds to repeated measurements. At 0.1 K, we observe a single signal peak corresponding to the tunneling of singlet ground state. As the temperature increases above 1 K, a second signal peak emerges at higher detuning, corresponding to triplet states tunneling from 11 to 02. Jumps in signal correspond to relaxation between singlet and triplet state. A charge shift occured between the low- and high-temperature measurements.
}
\label{fig:magneto}%
\end{figure}

To confirm the single occupancy regime of the three QDs in this state, we performed magnetospectrocopy on the ICTs  around the $\bigl(\begin{smallmatrix}
1&1\\ 1&
\end{smallmatrix} \bigr)$ charge state (Fig. \ref{fig:magneto}(a)). In particular, we focused on the $\bigl(\begin{smallmatrix}
1&1\\ 1&
\end{smallmatrix} \bigr)$ to $\bigl(\begin{smallmatrix}
0&1\\ 2&
\end{smallmatrix} \bigr)$ and the $\bigl(\begin{smallmatrix}
1&1\\ 1&
\end{smallmatrix} \bigr)$ to $\bigl(\begin{smallmatrix}
2&1\\ 0&
\end{smallmatrix} \bigr)$ ICTs.
Due to Pauli spin blockade, these transitions are assumed to present a spin-dependent quantum capacitance when two separated electrons form a triplet state.
Figure \ref{fig:magneto}(b) and (c) show how the signal on the ICTs changes as a function of magnetic field for these two transitions.
For both transitions, the main line attributed to a singlet state fades as the magnetic field increases. In these conditions, the triplet state $T^-_{11}$ becomes the ground state, giving rise to zero quantum capacitance and a null signal amplitude.
In parallel, a new line emerges at high magnetic field. This is strongly visible in Fig. \ref{fig:magneto}(b) for small detuning and more faintly in Fig. \ref{fig:magneto}(c) for larger detuning.
We attribute these transitions to tunneling between the $T^-_{11}$ and $T_{02}$ states (see Fig.\ref{fig:magneto}(d)) due to the presence of a low-lying excited state \cite{betz2015}.
The $T^-_{11}$ and $T_{02}$ states are probably valley states at \SI{100}{\micro\electronvolt} and \SI{220}{\micro\electronvolt}, respectively.
A simulation of the quantum capacitance of the first ICT taking into account the first low-lying valley state qualitatively and quantitatively reproduces the experimental data ( Fig. \ref{fig:magneto}(e)).

To further characterize our TQD system, we probed the ICT at various temperatures under zero field. The measurements presented in Fig. \ref{fig:magneto}(f) correspond to probing of the second ICT at 0.1 K and 1.2 K. At the lower temperature, signal is observed at one detuning, with the singlet state giving rise to measurable signal and jumps to the triplet state creating small blank pixels. At the higher temperature, a second signal line emerges at larger detuning. In this case, the signal is produced from the triplet state $T^-_{11}$ which tunnels to the $T_{02}$ state, and the blank pixels on the ICT correspond to occupancy of the singlet $S_{02}$.

\

\clearpage
\bibliography{biblio}

\newpage 
\clearpage

\end{document}